# Control of fluorescence in quantum emitter and metallic nanoshell hybrids for medical applications

Mahi R. Singh[1], Jiaohan Guo[1], José M. J. Cid[1,2], Jesús E. De Hoyos M.[2]

[1] Department of Physics and Astronomy,
The University of Western Ontario,
London N6A 3K7, Canada

[2] Posgrado, Facultad de Arquitectura y Diseño
Universidad Autónoma del Estado de México,
Toluca 50110, México

**Abstract**: We study the light emission from quantum emitter and double metallic nanoshell hybrid systems. Quantum emitters act as local sources which transmit their light efficiently due to a double nanoshell near field. The double nanoshell consists a dielectric core and two outer nanoshells. The first nanoshell is made of a metal and the second spacer nanoshell is made of a dielectric material or human serum albumin. We have calculated the fluorescence emission for a quantum emitter-double nanoshell hybrid when it is injected in an animal or human body. Surface plasmon polariton resonances in the double nanoshell are calculated using Maxwell's equations in the quasi-static approximation and the fluorescence emission is evaluated using the density matrix method in the presence of dipole-dipole interactions. We have compared our theory with two fluorescence experiments in hybrid systems in which the quantum emitter is Indocyanine Green or infrared fluorescent molecules. The outer spacer nanoshell of double metallic nanoshells consist of silica and human serum albumin with variable thicknesses. Our theory explains the enhancement of fluorescence spectra in both experiments. We find that the thickness of the spacer nanoshell layer increases the enhancement when the fluorescence decreases. The enhancement of the fluorescence depends on the type of quantum emitter, spacer layer and double nanoshell. We also found that the peak of the fluorescence spectrum can be shifted by changing the shape and size of the nanoshell. The fluorescence spectra can be switched from one peak to two peaks by removing the degeneracy of excitonic states in the quantum emitter. Hence using these properties one can use these hybrids as sensing and switching devices for applications in medicine.

# I: Introduction

Noble-metal nanoparticles are known to enhance emission rates of quantum emitters (QEs) significantly by decreasing their radiative lifetime and increasing their quantum yield [1-7]. The enhancement of the emission in QEs such as molecular fluorophores is a highly useful strategy for improving detection sensitivity and selectivity in many emerging applications, including DNA screening [5], single molecule detection [6] and image enhancement [7]. Consequently, designing and developing QE and metallic nanoparticle hybrids to enhance molecular fluorescence is of broad interest and applied importance. There is considerable interest to study hybrid systems made of biocompatible fluorescent molecules and metallic double nanoshells (DNSs) for biomedical imaging and for the detection of disease markers in the near-infrared wavelength region [8]. The penetration depth of near-infrared light is large in most biological media. It is found that these hybrids have large absorption coefficients and high quantum yields in the far-infrared region [9, 10]. Therefore, these hybrids can be used potentially for imaging deeply into the organs and soft tissues of living systems. They can also be used as agents for contrast enhancement and in physiological environments [11].

Recently Bardhan et al. [9] have fabricated a QE-DNS hybrid system to study the near-infrared fluorescence (FL). As QE they used Indocyanine Green (ICG), which is a fluorescent molecule. Currently it is the only FDA-approved and commercially available near-infrared-emitting dye. It is also used extensively as a fluorescent marker in clinical imaging applications [12] such as the diagnosis of cardiac and hepatic function [13], measurement of plasma volume and optical tomography [14]. However, ICG is a relatively weak fluorophore with a quantum yield of only 1.3% [15] and its toxicity limits the maximum concentration admissible for clinical uses. The DNS is fabricated from a silica core, a first Au nanoshell and a second silica nanoshell enclosing Au nanoshell. The second nanoshell is called the spacer layer. We denote this DNS as $SiO_2$-Au-$SiO_2$. Subsequently ICG molecules are deposited onto the spacer layer to complete the ICG-DNS hybrid. In this arrangement fluorescence enhancement of ICG molecules is observed as a function of the distance from the surface of the DNS hybrid. The distance between ICG molecules and the DNS surface is controlled by varying the thickness of the spacer layer. The enhancement of the absorption and FL emission of QEs will lead to significant improvements in the detection limits of the near infra-red fluorescence based imaging. For example, one can use these hybrids for detection of significantly smaller tumor volumes than it is currently possible.

Bardhan et al. [10] have also investigated the FL emission in IR800 molecules when they are placed on the spacer layer of a DNS. Here the DNS consists of a silica core, Au nanoshell and on outer spacer layer. In this case the space layer is fabricated from human serum albumin (HSA) and its thickness varies from 5 nm to 11 nm. We denote this DNS as HSA-Au-$SiO_2$. IR800 molecules, which act as QEs, are deposited on the HAS layer to fabricate the IR800-DNS hybrid. HSA is a large multidomain protein relevant to many physiological functions and has been conjugated to Au nanoparticles extensively for cell-targeting applications [16, 17]. It binds to Au by electrostatic attraction between the amine groups of HSA and the negative charge on the gold surface. The HSA acts as both a spacer layer as well as a linker for IR800 molecules. Bradhan et. al have measured the fluorescence enhancement of IR800 molecules. They found that the quantum yield of IR800 is enhanced as the thickness of the HSA layer is decreased.

In this paper, we study the fluorescence emission from QE-DNS systems in which QEs act as local sources and transmit their light efficiently in the presence of the DNS. The DNS exhibits unique and remarkably useful optical properties during excitation of surface plasmon polaritons (SPPs). SPP excitations emit significantly enhanced local fields near the surface of the metallic shell. This field interacts with the excitons of QEs and gives rise to fundamentally interesting physical phenomena for sensing and switching of the fluorescence emission. We have calculated SPPs for $SiO_2$-Au-$SiO_2$ and HAS-Au-$SiO_2$ double nanoshells using Maxwell's equations in the quasi-static approximation. We have found two SPP resonance energies due to the two interfaces present in these hybrids. It is found that the locations of SPP resonances in these hybrids can be manipulated by changing the size and shape of the metallic shell. We have also calculated the FL emission in QE-DNS hybrids using the density matrix method. The interaction between excitons in the QE and SPPs in the DNS was included in the FL evaluation. This interaction is also called the dipole-dipole interaction (DDI). When this hybrid is injected in an animal or human body, it is surrounded by biological cells such as cancer cells. The effect of biological cells has also been included in the FL and SPPs evaluations.

Finally, we have compared our theory with the fluorescence emission of IGL-DNS hybrid [9]. Here the DNS is made from silica core, Au-nanoshell and the silica spacer layer. Then IGL molecules are deposited on the surface of the Au-DNS. We have also compared our theory with the second hybrid IR800-DNS [10]. In this case the DNS is fabricated from a silica core, Au-nanoshell and a HAS space layer. The IR800 molecules are deposited on the HSA layer. In our calculations, we have considered that IR800 and IGL molecules have two degenerate excitons which are interacting with SPPs of the DNS. Good agreement between our theory and the experimental literature data is found when the distance between the QE and DNS is varied. We have also calculated the fluorescence for a QE which has two non-degenerate excitons. We found that the FL spectrum splits from one peak to two peaks in the presence of exciton-SPP coupling (i.e. DDI). These interesting findings may be useful in the fabrication of nanosensors and nanoswitches for applications in medicine.

## II: Polarizability and surface plasmon polaritons of double metallic nanoshells

In this section, we calculate the polarizability and SPP resonance frequencies. For this we model the DNS as consisting of a dielectric core, a metallic nanoshell and an outer spacer layer nanoshell. The DNS has a spheroidal shape. QEs are deposited on the spacer layer to complete the QE-DNS hybrid. We denote the core, the metallic nanoshell and the spacer layer as 1, 2 and 3, respectively. Dielectric constants for the core, metallic nanoshell and the spacer layer are denoted as $\epsilon_1$, $\epsilon_2$ and $\epsilon_3$, respectively. The hybrid is assumed to be injected into a human or animal cell. This means that the hybrid is surrounded by bio-cells and the dielectric constant of bio-cells is denoted as $\epsilon_b$. We denote the volumes of the core, metallic shell and core and DNS as $V_1$, $V_2$ and $V_3$, respectively. A schematic diagram of the DNS is depicted in fig. 1.

At the interface between the metallic shell and the core surface plasmon polaritons (SPPs) are present. Similarity, SPPs are also found at the interface between the spacer layer and the metallic shell. SPP resonance frequencies are calculated from the polarizability of the DNS. Hence we calculate the polarizability of the DNS as follows. A probe laser light with frequency ω, wavelength λ and electric field $E_p$ is applied to the DNS. The typical size of the DNS is less than one hundred nanometers. The wavelength of light in the visible region is of the order of several hundred nanometres. This means that the size of the DNS is much smaller than the wavelength of light. Hence, we consider that the amplitude of electric field is constant over the DNS. This is known as the quasi-static approximation [18-20]. Solving Maxwell's equations in this approximation and using the transfer matrix method, the polarizability $α_{NS}$ of the DNS is found as

$$\alpha_{NS} = \epsilon_0 \epsilon_b V_3 \zeta_{NS} \qquad (1)$$

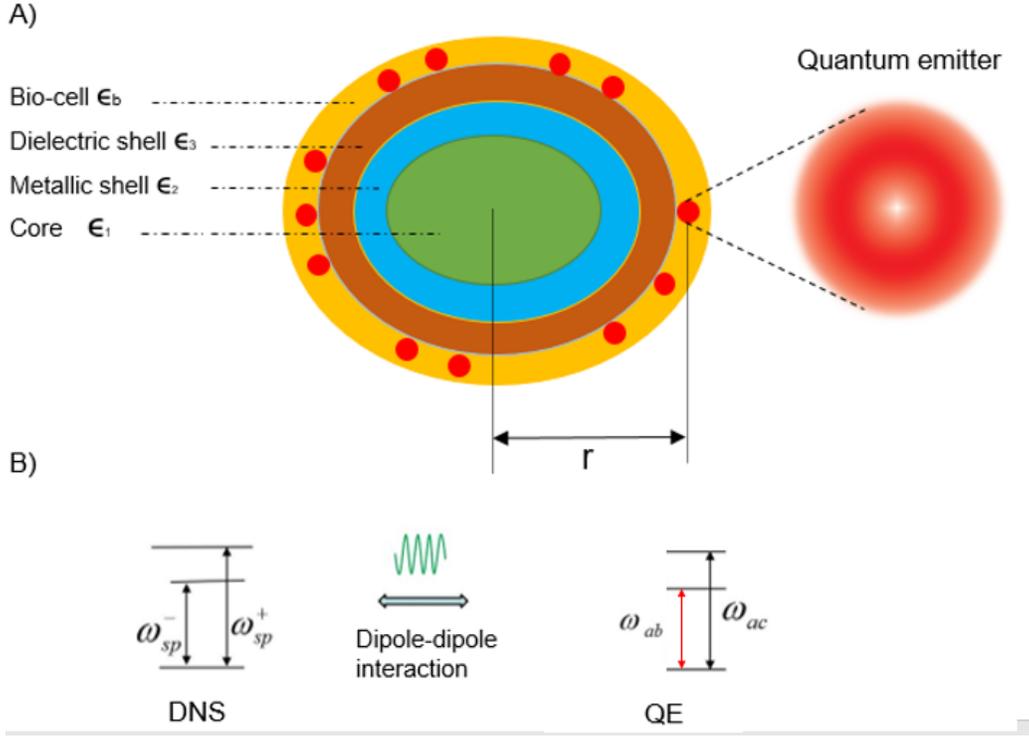

Fig. 1: (A) Schematic diagram of the double nanoshell. The dielectric constants of the inner core material, metallic nanoshell and the spacer layer are denoted as $\epsilon_1$, $\epsilon_2$ and $\epsilon_3$ respectively. The DNS is doped into a bio-cell with dielectric constant $\epsilon_b$. QEs are deposited on the surface of the DNS. The distance between the DNS and QE is denoted as $r$. (B) The energy levels of the DNS and WE are plotted. The QE has two excitonic states $\omega_{ab}$ and $\omega_{ac}$, the DNS has two localised SPP states $\omega_{sp}^+$ and $\omega_{sp}^-$.

where $\epsilon_0$ is the vacuum permittivity and $\varsigma_{NS}$ is called the polarizability factor which has the following form

$$\varsigma_{NS} = \frac{[\epsilon_{32} - \epsilon_b]}{3\xi_3 \epsilon_{32} + (3 - 3\xi_3)\epsilon_b} \qquad (2)$$

Parameters appearing in eqn. (2) are calculated as

$$\epsilon_{32} = \epsilon_3 \left[\frac{1 + 2r_{32}\varsigma_{32}}{1 - r_{32}\varsigma_{32}}\right] \qquad (3a)$$

$$\epsilon_{21} = \epsilon_2 \left[\frac{1 + 2r_{21}\varsigma_{21}}{1 - r_{21}\varsigma_{21}}\right] \qquad (3b)$$

$$\varsigma_{32} = \frac{[\epsilon_{21} - \epsilon_3]}{3\xi_2 \epsilon_{21} + (3 - 3\xi_2)\epsilon_3} \qquad (4a)$$

$$\varsigma_{21} = \frac{[\epsilon_1 - \epsilon_2]}{3\xi_1 \epsilon_1 + (3 - 3\xi_1)\epsilon_2} \tag{4b}$$

with $r_{32} = V_2 / V_3$ and $r_{21} = V_1 / V_2$. Parameter $\xi_i$ is called the depolarization factor and depends on the shape of the DNS. Its expression is obtained as [19]

$$\xi_i = \frac{V_i}{2} \int_0^\infty \frac{ds}{(s + L_{ix}^2)^{3/2}(s + L_{iy}^2)^{1/2}(s + L_{iz}^2)^{1/2}} \tag{5}$$

where $L_{ix}$, $L_{iy}$ and $L_{iz}$ with i =1, 2, 3 are the length, width and thickness of the $i^{th}$ material, respectively. Note that it is a dimensionless quantity.

We can also calculate the polarizability for the spherically shaped DNS from eqns. (1-5). We consider that the radii of the core, metallic shell and DNS are denoted as $R_1$, $R_2$ and $R_3$, respectively. The expression of the polarizability is found as

$$\alpha_{NS} = 4\pi \epsilon_0 \epsilon_b R_3^3 \varsigma_{NS} \tag{6}$$

where $\varsigma_{NS}$ is calculated as

$$\varsigma_{NS} = \frac{[\epsilon_{32} - \epsilon_b]}{\epsilon_{32} + 2\epsilon_b} \tag{7}$$

$$\epsilon_{32} = \epsilon_3 \left[\frac{R_3 + 2R_2\varsigma_{32}}{R_3 - R_2\varsigma_{32}}\right], \quad \varsigma_{32} = \frac{[\epsilon_{21} - \epsilon_3]}{\epsilon_{21} + 2\epsilon_3}$$

$$\epsilon_{21} = \epsilon_2 \left[\frac{R_2 + 2R_1\varsigma_{21}}{R_2 - R_1\varsigma_{21}}\right], \quad \varsigma_{21} = \frac{[\epsilon_1 - \epsilon_2]}{\epsilon_1 + 2\epsilon_2} \tag{8}$$

Note that the above expression depends on the radii of the DNS. Hence the polarizability depends on the radii and the dielectric constants of the constituents of the DNS.

We use the Drude model for the dielectric function of the metallic shell in the form:

$$\epsilon_2 = \epsilon_\infty - \frac{\omega_p^2}{\omega^2 + i\omega\gamma_m} \tag{9}$$

where $\omega_p$ is the plasmon frequency of the metal and $\epsilon_\infty$ is its relative permittivity at very large energies $(\omega \gg \omega_p)$. $\gamma_m$ is called the decay rate and represents the thermal energy loss in the metallic shell.

The singularity in the polarizability gives the locations of SPP frequencies and can be found by equating the denominator of the polarizability factor to zero. The DNS structure has two surface plasmon polaritons propagating within the two interfaces. Let us denote the SPP frequency (wavelength) at the interface between the metallic shell and the core as $\omega_{SP}^-$ ($\lambda_{SP}^-$), and between the metallic shell and the spacer layer as $\omega_{SP}^-$ ($\lambda_{SP}^-$). Note that the polarizability for the spheroidal given in eqn. (2) depends on $r_{32}$, $r_{21}$ and $\xi_i$. This means by adjusting the shape of the DNS, the SPP resonance wavelengths can be made larger than the size of the DNS. These resonances can be varied from UV to IR wavelengths of light.

## III: Fluorescence and exciton-SSP interaction

In this section, we calculate the FL emission spectrum for the QE in the QE-DNS hybrid using the density matrix method in the presence of exciton-SPP interaction. The QE has excitons (electron-hole pairs) and the DNS contains SPPs. These interact with each other via dipole-dipole interactions (DDI). We consider that the QE has three-levels denoted as $|a\rangle$, $|b\rangle$ and $|c\rangle$. It has two excitons with frequencies (wavelengths) $\omega_{ab}$ ($\lambda_{ab}$) and $\omega_{ac}$ ($\lambda_{ac}$), which are due to the transitions $|a\rangle \to |b\rangle$ and $|a\rangle \to |c\rangle$, respectively. A schematic diagram of the QE is shown in fig. 1.

We apply a probe field to monitor the FL emission due to the transitions $|a\rangle \to |b\rangle$ and $|a\rangle \to |c\rangle$. Due to the probe field an induced polarization is created in the DNS as $P_{SP} = \alpha_{SP} E_P$ which in turn produces a SPP dipole electric field ($E_{SP}$) at a distance $r$ from the centre of the DNS. It is found that [20, 21]

$$E_{SP} = \frac{g_l P_{SP}}{4\pi \epsilon_0 \epsilon_b r^3} = \frac{g_l V_3 \zeta_{NS}}{4\pi r^3} E_p \qquad (10)$$

The constant $g_l$ is called the polarization parameter and has the values $g_l = -1$ and $g_l = 2$ for $P_{SP} \| E_p$ and $P_{SP} \perp E_p$, respectively. Note that the SPP electric field depends on the polarizability factor $\zeta_{NS}$. The polarizability factor has its largest values at $\omega = \omega_{SP}^+$ and $\omega = \omega_{SP}^-$, and they are denoted as $\zeta_{SP}^+$ and $\zeta_{SP}^-$, respectively. For other $\omega$ values the polarizability factor is neglected since its effect in the calculation of the FL will be negligible. Therefore, the SPP electric field given in eqn. (10) can be expressed as $E_{SP} = E_{SP}^+ + E_{SP}^-$ where $E_{SP}^+$ and $E_{SP}^-$ correspond to $\zeta_{SP}^+$ and $\zeta_{SP}^-$, respectively. Note that the SPP electric field depends on the distance r from the DNS.

Let us first calculate the interaction Hamiltonian between the QE and the DNS. The probe and SSP electric fields are falling on the QE. Hence the total field seen by the QE can be expressed as

$$E_{QE} = E_P + E_{SP}^+ + E_{SP}^- \qquad (11)$$

The probe and SPP fields induce dipoles in the QE due to exciton transitions $|a\rangle \to |b\rangle$ and $|a\rangle \to |c\rangle$. Hence these induced dipoles interact with the applied electric fields and this interaction is called the exciton-SPP interaction. The Hamiltonian for this interaction is expressed in the dipole and rotating wave approximation as

$$H_{ex-spp} = \sum_{n=a,b,c} \hbar\omega_n \sigma_{nn} - \sum_{n=b,c} \hbar\Omega_n \sigma_{na}^+ - \sum_{n=b,c} \hbar\Pi_n \sigma_{na}^+ + h.c.$$

$$\Omega_n = \frac{\mu_{na} E_P}{2\hbar}, \qquad \Pi_b = \frac{g_l V_3 \zeta_{NS}^-}{4\pi r^3}\Omega_b, \qquad \Pi_c = \frac{g_l V_3 \zeta_{NS}^+}{4\pi r^3}\Omega_c \qquad (12)$$

where h.c. stands for the Hermitian conjugate and $\sigma_{na}^+ = |n\rangle\langle a|$ with $|n\rangle = |b\rangle, |c\rangle$ is the exciton creation operator. Here $\Omega_{na}$ is called the Rabi frequency which is associated with the transition $|n\rangle \to |b\rangle$. The first term in Eq. (12) is the Hamiltonian of the QE. The second term is the interaction between the exciton and the external probe field $E_p$. The third term is due to the interaction of dipoles of the QE with the dipoles of the DNS (i.e. SPPs). This term is also called the DDI because induced dipoles in QE and induced dipoles in the DNS interact with each other. We consider that the SPP frequency $\omega_{SP}^+$ lies close to exciton frequency $\omega_{ac}$. Similarly, the SPP frequency $\omega_{SP}^-$ lies close to exciton frequency $\omega_{ab}$.

We consider that excitons in the QE decay from excited states $|b\rangle$ and $|c\rangle$ to the ground state $|a\rangle$ and lose energy due spontaneous emission. This is called the radiative decay rate. It is also considered that excitons decay due to the exciton-SPP interaction and lose energy to the DNS. This is known as the nonradiative decay rate. The decay interaction Hamiltonian can be written in the second quantized notation using the rotating wave approximation as

$$H_{int} = -\sum_k \sum_{n=b,c} V_n^r(\omega_k) a_k \sigma_{na}^\dagger + \sum_\beta \sum_{n=b,c} V_n^{nr}(\omega_\beta) a_\beta \sigma_{na}^\dagger + h.c. \qquad (13)$$

where $V_{na}^r$ and $V_{na}^{nr}$ are coupling constants for the radiative and nonradiative interactions, respectively. They are found as

$$V_{na}^r = i\left(\frac{\varepsilon_k}{2\epsilon_0 \pi V_{QE}}\right)^{1/2} (\boldsymbol{\mu}_{na} \cdot \mathbf{e}_k) \qquad (14)$$

$$V_n^{nr} = i\left(\frac{\varepsilon_\beta}{2\epsilon_0 \pi V_{QE}} \frac{g_l V_3 \zeta_{NS}}{4\pi r^3}\right)^{1/2} (\boldsymbol{\mu}_{na} \cdot \mathbf{e}_k), \qquad (15)$$

where $\mu_{na}$ is the induced dipole moment due to the transition $|n\rangle \to |b\rangle$, operator $a_k$ is the photon annihilation operator, operator $a_\beta$ is the SPP annihilation operator and $V_{QE}$ is the volume of the QE.

Using the master equation for the density matrix [20-24] and Eqs. (12, 13) for the Hamiltonian of the system, we obtained the following equations of motion for the density matrix elements

$$\frac{d\rho_{cc}}{dt} = -\gamma_c \rho_{cc} - i(\Omega_c + \Pi_c)\rho_{ac} - i(\Omega_c + \Pi_c)^* \rho_{ca} \qquad (16a)$$

$$\frac{d\rho_{bb}}{dt} = -\gamma_b \rho_{bb} + i(\Omega_b + \Pi_b)\rho_{ab} - i(\Omega_b + \Pi_b)^* \rho_{ba} \qquad (16b)$$

$$\frac{d\rho_{ca}}{dt} = -\Xi_{ca}\rho_{ca} - i(\Omega_c + \Pi_c)(\rho_{cc} - \rho_{aa}) - i(\Omega_b + \Pi_b)\rho_{cb} \tag{16c}$$

$$\frac{d\rho_{ba}}{dt} = -\Xi_{ba}\rho_{ba} - i(\Omega_b + \Pi_b)(\rho_{bb} - \rho_{aa}) - i(\Omega_c + \Pi_c)\rho_{bc} \tag{16d}$$

$$\frac{d\rho_{cb}}{dt} = -\Xi_{cb}\rho_{cb} + i(\Omega_c + \Pi_c)\rho_{ab} - i(\Omega_b + \Pi_b)^* \rho_{ca} \tag{16e}$$

Parameters appearing in eqn. (16) are found as

$$\Xi_{ca} = \frac{\gamma_c}{2} + i\delta_c$$

$$\Xi_{ba} = \frac{\gamma_b}{2} + i\delta_b \tag{17}$$

$$\Xi_{cb} = \frac{\gamma_b + \gamma_c}{2} + i(\delta_c - \delta_b)$$

Here $\delta_b = \omega_{ab} - \omega$ and $\delta_c = \omega_{ac} - \omega$ are called probe field detunings. Physical quantities $\gamma_b$ and $\gamma_c$ are decay rates of levels $|b\rangle$ and $|c\rangle$, respectively, and are found as

$$\gamma_b = \gamma_b^r + \gamma_b^{nr}, \quad \gamma_c = \gamma_c^r + \gamma_c^{nr} \tag{18}$$

Here $\gamma_n^r$ and $\gamma_n^{nr}$ with $n = b, c$ are radiative and nonradiative decay rates, respectively. They are found as

$$\gamma_b^r = \frac{\mu_{ba}^2 \varepsilon_{ba}^3}{3\pi \epsilon_0 \hbar^4 c^3}, \quad \gamma_b^{nr} = \gamma_b^r \left(\frac{g_l V_3 \zeta_{NS}^-}{4\pi r^3}\right)$$

$$\gamma_c^r = \frac{\mu_{ca}^2 \varepsilon_{ca}^3}{3\pi \epsilon_0 \hbar^4 c^3}, \quad \gamma_c^{nr} = \gamma_c^r \left(\frac{g_l V_3 \zeta_{NS}^+}{4\pi r^3}\right) \tag{19}$$

Note that the radiative decay rate depends on the SSP frequency and the distance between the QE and the surface of the DNS.

Finally, we calculate the expression of the FL emitted from the QE. The method for this has been developed for nanoparticles in reference [21]. Here we use this method to develop a theory for the present paper. Following the method of reference [21] we have evaluated the FL expression for the QE-DNS hybrid system as

$$L_{QE} = Q_{QE} W_{QE} \tag{20}$$

where

$$L_{QE} = Q_{QE}W_{QE} \qquad Q_{QE} = \frac{\gamma_c^r}{\gamma_c^r + \gamma_c^{nr}}$$

$$W_{QE} = \text{Im}\left(\frac{\mu_{ba}\omega_{ba}\rho_{ba} + \mu_{ca}\omega_{ca}\rho_{ca}}{2E_p}\right)|E_{QE}|^2$$

In the above equation $Q_{QE}$ is called the FL efficiency factor and $W_{QE}$ is the power emitted by the QE. To solve eqn. (20) we need to evaluate $\rho_{ba}$ and $\rho_{ca}$ form eqn. (16) using numerical methods.

However, we can obtain an analytical expression for $\rho_{ba}$ and $\rho_{ca}$ in the steady state. We consider that the ground state population of the excitons ($\rho_{aa}$) is much larger than the population of the excited states ($\rho_{bb}, \rho_{cc}$). This gives the condition $\rho_{aa} \gg \rho_{bb}$ and $\rho_{aa} \gg \rho_{cc}$. These density matrix elements satisfy the condition $\rho_{aa} + \rho_{bb} + \rho_{cc} = 1$. For simplicity, we also consider that $\mu_{ba} = \mu_{ca} = \mu$. This gives us $\Omega_b = \Omega_c = \Omega$ and $\gamma_b = \gamma_c = \gamma$. Solving eqn. (16) for $\rho_{ba}$ and $\rho_{ca}$ in the steady state and substituting in eqn. (20), we get

$$L_{QE} = L_{QE}^b \text{Im}\left(\frac{4(1+\Lambda_b)}{(\delta_b/\gamma - i)}\right)|1+\Lambda_b|^2 + L_{QE}^c \text{Im}\left(\frac{4(1+\Lambda_c)}{(\delta_c/\gamma - i)}\right)|1+\Lambda_c|^2 \qquad (21)$$

where

$$L_{QE}^b = \hbar\omega_{ba}\gamma(\Omega/\gamma)^2\left(\frac{\gamma_b^r}{\gamma_b^r + \gamma_b^{nr}}\right), \qquad L_{QE}^c = \hbar\omega_{ca}\gamma(\Omega/\gamma)^2\left(\frac{\gamma_c^r}{\gamma_c^r + \gamma_c^{nr}}\right)$$

$$\Lambda_b = g_l \in_b \left(\frac{V_3}{r^3}\right)\varsigma_{NS}^-, \qquad \Lambda_c = g_l \in_b \left(\frac{V_3}{r^3}\right)\varsigma_{NS}^+ \qquad (22)$$

Next, we calculate the FL when the two excitons have degenerate excitonic states. This means we have $\omega_{ca} \approx \omega_{ba}$. In this case the FL expression given in eqn. (21) reduces to

$$L_{QE} = L_{QE}^0 \text{Im}\left(\frac{4(1+\Lambda_b+\Lambda_c)(\delta_k/\gamma - i)}{(\delta_k/\gamma - i)^2 - \omega_{cb}^2/\gamma^2}\right)|1+\Lambda_b+\Lambda_c|^2 \qquad (23)$$

Where

$$L_{QE}^0 = \hbar\omega_{ba}\gamma(\Omega/\gamma)^2\left(\frac{\gamma_c^r}{\gamma_c^r + \gamma_c^{nr}}\right)$$

In eqn. (23) $\delta_k$ is called the probe detuning and is defined as $\delta_k = \omega_{ca} + \omega_{ba} - 2\omega$.

Note that the FL depends on $\Lambda$ which in turn is inversely proportional to $r^3$. This means the FL depends on the DDI between QE and DNS. For smaller distances the DDI term has large values. It also depends on the polarizability factor $\varsigma_{NS}^\pm$ which has large values at the SPP resonance energies.

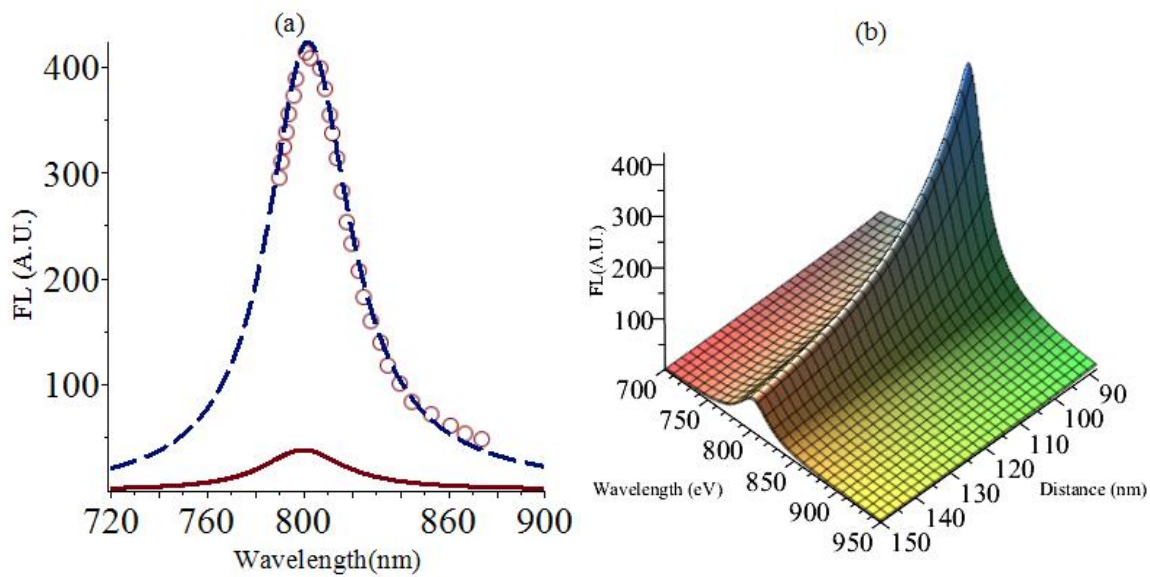

Fig. 2: (a)Plot of the FL (A.U.) as a function of wavelength (nm) for the IR800-Au-DNS. A.U. means arbitrary units. The core of the Au-nanoshell is made of silica (SiO$_2$) with a radius of 63 nm and the Au nanoshell with a radius of 76 nm. The Au-nanoshell is coated with human serum albumin (HSA) with thickness 8 nm. The open circles denote the experimental data. The solid curve is the theoretical result when the exciton wavelength and SSP wavelength are not in resonance ($\lambda_{sp} \neq \lambda_{ab}$). Similarly, the dashed curve is plotted when the exciton wavelength and SSP wavelength are in resonance ($\lambda_{sp} = \lambda_{ab}$). Other parameters are taken as $\hbar\omega_p = 9$ eV and $\epsilon_\infty = 8$, $\gamma_m = 0.3$ eV. The dielectric constant for the silica and HAS are taken as $\epsilon_1 = 1.3$ and $\epsilon_3 = 1.35$, respectively. (b) Three-dimensional plot of the FL (A.U.) as a function of wavelength (nm) and thickness of the HSA layer.

## IV: Results and Discussion

We have developed a theory for the FL in eqn. (23) for a QE-DNS hybrid and this equation is used to explain the experimental data of the FL emission presented in references [9, 10]

Recently QE-DNS hybrids have been fabricated by several groups [9, 10] for medical purposes. First, we compare our theory with the work of Bardhan et al. [9]. These authors have fabricated a IR800-Au-DNS hybrid system. The core of the Au-nanoshell is made of silica (SiO2) with a radius of 63 nm. It is coated with a Au shell to a radius of 76 nm. Finally, the Au-nanoshell is coated with HAS, which serves as a spacer layer to construct the DNS. Subsequently the hybrid system was fabricated by absorbing IR800 fluorophore molecules onto the DNS electrostatically, to get IR800-DNS hybrid. They have studied the fluorescence enhancement of the IR800 fluorophore as a function of distance from the surface of DNS. They found FL enhancement in the near-infrared region in IR800-fluorophore in the presence of the DNS by varying the thickness of the HAS spacer layer. Their measurements reveal that the quantum yield of IR800 is enhanced from 7% as an isolated fluorophore to 86% in the IR800 in the IR800-DNS hybrid. This dramatic fluorescence enhancement can be used for contrast enhancement in fluorescence-based bioimaging.

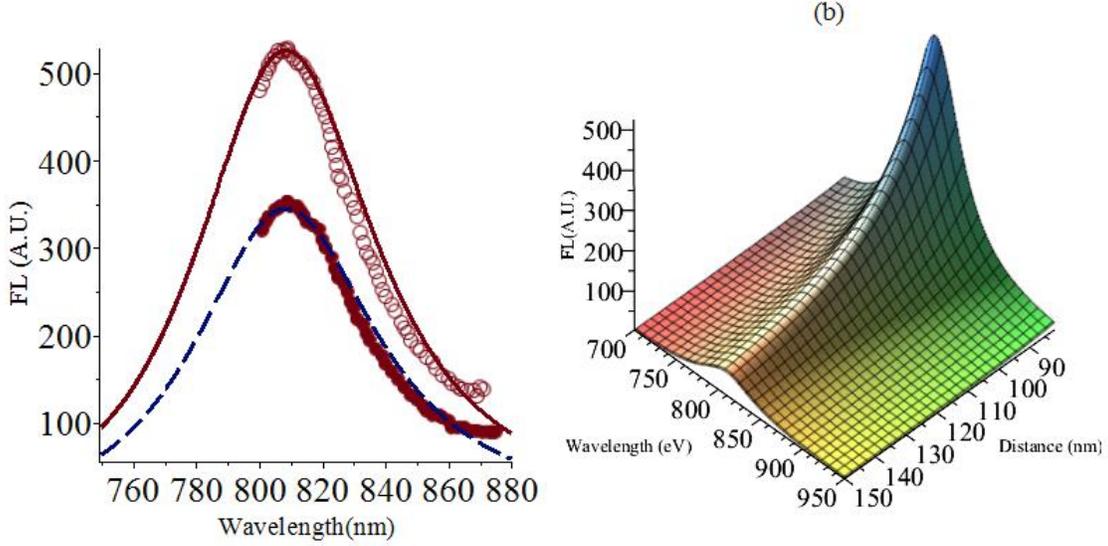

Fig. 3: Plot of the FL (A.U.) as a function of wavelength (nm). A.U. stands for arbitrary units. The open circles and solid diamonds denote the experimental data when the thickness of the silica is 7nm and 13nm, respectively. The solid and dashed curves are theoretical results when the thickness of the silica is 7nm and 13nm, respectively. The radii of the core and metallic shell are taken as 60 nm and 72 nm, respectively. Other parameters are taken as $\hbar\omega_p = 9$eV and $f_{ss} = 8$, $\gamma_m = 0.7$ eV. The dielectric constant for the silica is taken as $\epsilon_1 = \epsilon_3 = 1.3$.

Now we compare our theory with above experimental data [9]. We consider that the IR800 fluorophore molecules acts as QE and have two excitonic states with wavelength $\lambda_{ab}$ and $\lambda_{ac}$ and both states are degenerate ($\lambda_{ab} = \lambda_{ac}$). The SPP wavelength of the Au-DNS is calculated as $\lambda_{sp} = 805$ nm to agree with the experimental value. The FL theoretical and experimental data are plotted as a function of the wavelength (nm) in fig. 2. The open circles denote the experimental data when thickness of the HSA is 8 nm. The solid curve is plotted when the exciton wavelength and SSP wavelength are not in resonance ($\lambda_{sp} \neq \lambda_{ab}$). The dotted curve is the theoretical result when the exciton wavelength and the SSP wavelength are in resonance with the exciton wavelength ($\lambda_{sp} = \lambda_{ab}$). There is a dramatic enhancement in the FL emission when the exciton wavelength and the SPP wavelength are in resonance. Note that a good agreement between theory and experiment is found. We have also plotted a three-dimensional figure for the FL spectrum as a function of the thickness of the HAS layer and the wavelength. One can see that the FL enhancement is increased as the thickness of the HAS layer is decreased. This is consistent with the experimental results of reference [9].

Further we investigate experimental work of Bardhan et al. [10] where the authors have fabricated the ICG-DNS hybrid. Here ICG molecules act as a QE. The core of the DNS is made of silica (SiO2) with a radius of 60nm. The core is coated with an Au shell and its radius is 72nm. Then the Au-shell is coated with varying thicknesses of silica spacer layers to complete the DNS. Consequently, the hybrid system was fabricated by absorbing ICG molecules onto the spacer layer electrostatically to complete ICG-DNS hybrid. The silica spacer layer can be manipulated to change the distance between the ICG and Au-nanoshell. Authors have investigated the FL enhancement of ICG molecules as a function of distance from the surface of the Au-nanoshell. The distance between ICG molecules and the Au-nanoshell surface is controlled by varying the thickness of the outer silica shell. A fluorescence enhancement is observed when ICG molecules are spaced closer to the Au-nanoshell surface. The enhancement decreases with increasing distance of the molecule from the Au-nanoshell surface.

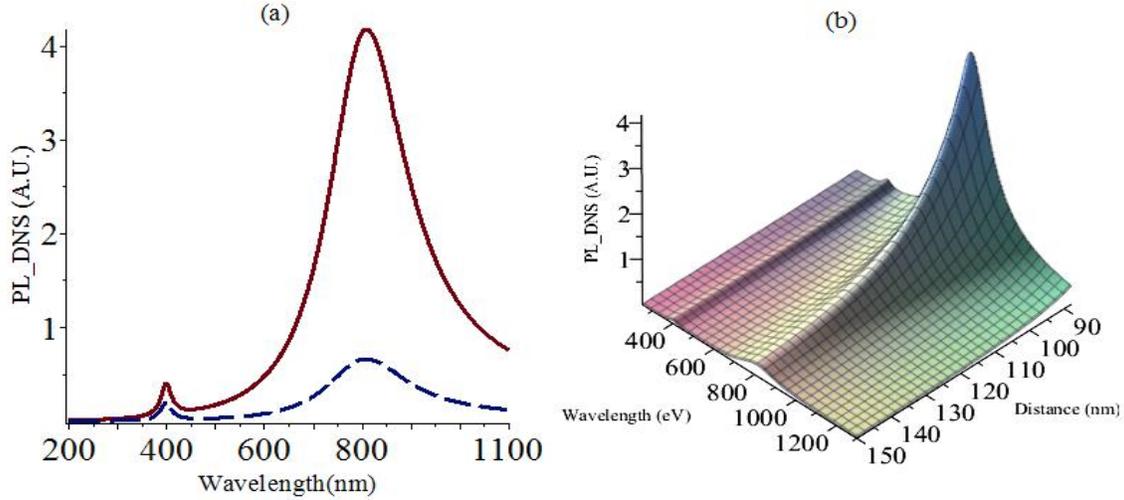

Fig. 4: **(a)** Plot of the FL (A.U.) as a function of wavelength (nm). The solid and dotted curves are theoretical results for when r = 86 nm and 112 nm, respectively. **(b)** Three dimensional plot of the FL (A.U.) as a function of wavelength (nm) and distance between the QE and the DNS (r). The radius of the core and metallic shell are taken as 60 nm and 72 nm, respectively. Other parameters are taken as $\hbar\omega_p$ = 9eV and $g_{ii}$ = 8, $\gamma_m$ = 0.7 eV. The dielectric constant for the silica is taken as $\epsilon_1 = \epsilon_3 = 1.3$.

In fig. 3 we compare our theory with the experimental data of the ICG-DNS hybrid [10]. We consider that ICG molecules act as QE and have two degenerate excitonic states with wavelength $\lambda_{ab}$. The SPP wavelength of the Au-DNS is calculated as $\lambda_{sp}$ = 810nm. The theoretical and experimental date are plotted in fig. 3a. Open circles and solid diamonds denote the experimental data when the thicknesses of the silica spacer layer are 7nm and 13nm, respectively. The solid and dotted curves are theoretical results. In plotting theoretical curves, we have considered that the exciton wavelength and SSP wavelength are in resonance with each other ($\lambda_{sp} = \lambda_{ab}$). Note that a good agreement between theory and experiment is found when the exciton and SSP wavelength are in resonance. If this condition were not satisfied, then our theory would not agree with experiments. We have also plotted a three-dimensional figure of the FL by varying the spacer thickness. Note that the enhancement decreases as the distance between ICG and DNS increases. This finding is consistent with experimental data of reference [10].

We also study the FL for the IR800-DNS hybrid when the QE is in the nondegenerate state. This means two excitons frequencies are different ($\lambda_{ab} \neq \lambda_{ac}$). In plotting theoretical curves, we have considered that the exciton wavelength $\lambda_{ac}$ and the SSP wavelength ($\lambda_{SP}^+$) are in resonance with each other. Similarly, the second exciton wavelength $\lambda_{ab}$ and the second SSP wavelength ($\lambda_{SP}^-$) are also in resonance. The results are plotted in Fig. 4a and 4b. In fig 4a the FL spectrum is plotted as a function of wavelength. The solid and dashed curves are plotted when the distance between IR800 and DNS are r = 86 nm and r =112 nm, respectively. Fig 4b contains the three-dimensional plot of the FL as a function of wavelength and distance between the QE and DNS. It is found that the FL spectrum has two peaks. The peak related to $\lambda_{ab}$ is weaker than that of $\lambda_{ac}$. This is because the exciton-DDI coupling is weaker for the $\lambda_{ab}$-peak compared to that of the $\lambda_{ac}$-peak. Note that enhancement of the FL intensity increases as the distance between the QE and DNS decreases. This finding is consistent with experimental results of references [9, 10]. Similar results are also found for the IR800-DNS hybrid. Note that the FL enhancement in the second peak is huge compared to the first peak.

Our theory is valid for any dielectric material used as core or spacer layer for the fabrication of DNS such as Silica, $Y_2O_3$ and $YPO_4$. The effect of these materials in our theory is represented by their corresponding dielectric constants (i.e. see eqns. (1-9)). The dielectric constant of these materials plays an important role in our theory. According to our theory, the locations of the SPP resonance frequencies and the intensity of the SPPs electric field depend on the dielectric constant of the materials used in the DNS fabrication (see eqns. (6, 10)).

Let us consider that QEs (ICG/IR800) are placed on the metal surface instead of the spacer layer of the DNS. This means we have a single nanoshell which is made of a metallic shell and a silica core. The intensity of the FL emitted from QEs will be stronger than when QEs are capped on the surface of the DNS. This is because surface plasmons polaritons are created at the metallic surface where the electric field intensity has the largest value. The dipole-dipole interaction between QEs and a single metallic shell will also be enhanced since the distance between the QE and the metallic surface is smaller in the absence of the spacer layer. We have shown that DDI depends on the distance from the metal surface to the QE. We have also found that the DDI is responsible for the enhancement of the FL in QEs.

In summary, our theory explains the enhancement of the FL spectra for IR800-DNS and ICG-DNS hybrids. We showed that the enhancement in the FL is decreased as the thickness of the spacer layer is increased. The enhancement in the FL emission depends on the type of the QE, spacer layer and the DNS. We also found that the peak of the FL can be shifted by changing the shape and size of the DNS. It is found that the FL spectra can be switched from one peak to two peaks by removing the degeneracy of excitonic states in the QE. Hence using these properties one can use these hybrids as sensing and switching devices for applications in medicine.

## V: Conclusions

Light emission from QE-DNS hybrids was studied. The DNS is made of a dielectric core coated with a thin layer of metal and then further coated by a spacer layer of dielectric material. Subsequently QEs are deposited on the surface of the spacer layer to fabricate the QE-DNS hybrid. When this hybrid is injected in animal or human tissue it is surrounded by biological cells such as cancer cells. Surface plasmon polariton resonances in the DNS are calculated using Maxwell's equations in the quasi-static approximation and the FL is evaluated using the density matrix method in the presence of the DDI. It is found that locations of SPP resonances in DNS can be manipulated by changing the size and shape of the metallic shell, core and spacer layer. We have compared our theory of the FL emission with experimental results for the IGL-DNS hybrid where the DNS is made of a silica core, Au shell and the silica spacer layer. We have also compared our theory with experimental results for the IR800-DNS hybrid. In this case, the DNS consists of a silica core, Au shell and an HSA spacer layer. Good agreement between theory and experiment is found when the exciton energy and the SSP energy are in resonance. We have also calculated the FL for the nondegenerate QE and found that the FL spectrum splits from one peak to two peaks. These interesting findings may be useful in the fabrication of nanosensors and nanoswitches for applications in medicine.

## Acknowledgement

One of the authors (MRS) is thankful to the Natural Sciences and Engineering Research Council of Canada (NSERC) for their research grant and José M. J. Cid who appreciative to Universidad Autónoma del Estado de México for awarding scholarship. Authors are thankful to Professor M. Zinke-Allmang for editing the paper.

## References

1. R. D. Artuso, G. W. Bryant, Phys. Rev. B82, 195419 (2010).
2. S. M. Sadeghi, L. Deng, X. Li, W. P. Huang, Nanotechnology. 20, 365401 (2009).
3. J. D. Cox, M. R. Singh, G. Gumbs, M. A. Anton, F. Carreno, Phys. Rev. B86, 125452 (2012).
4. M. Singh, C. Manda, S. Balakrishnan, Nanotechnology. (submitted 2016).
5. A. C. Pease, D. Solas, E. J. Sullivan, M. T. Cronin, C. P. Holmes, S. P. A. Fodor, Proc. Natl. Acad. Sci. USA. 91, 5022 (1994).
6. S. Weiss, Science. 283, 1676 (1999).
7. A. W. Weinberger, B. Kirchhof, B. E. Mazinani, N. F. Schrage, Graefe's Arch. Clin. Exp. Ophthalmol. 239, 388 (2001).
8. K. E. Adams, S. Ke, S. Kwon, F. Liang, Z. Fan, Y. Lu, K. Hirschi, M. E. Mawad, M. A. Barry, E. M. Sevick-Muraca, J. Biomed. Opt. 12(2), 024017 (2007).
9. R. Bardhan, N. K. Grady, J. R. Cole, A. Joshi, N. J. Halas, ACS Nano. 3, 744 (2009).
10. R. Bardhan, N. K. Grady, N. J. Halas, Small. 4, No. 10, 1716 (2008).
11. J. P. Houston, S. Ke, W. Wang, C. Li, E. M. Sevick-Muraca, J. Biomed. Opt. 10, 0540101 (2005).
12. T. Tsubono, S. Todo, N. Jabbour, A. Mizoe, V. Warty, A. J. Demetris, T. E. Starzyl, Hepatology. 24, 1165 (1996).
13. S. G. Sakka, K. Reinhart, K. Wegscheider, A. Meier-Hellmann, Chest. 121, 559 (2002).
14. V. Ntziachristos, A. G. Yodh, M. Schnall, B. Chance, Proc. Natl. Acad. Sci. USA. 97, 2767 (2000).
15. R. C. Benson, H. A. Kues, Phys. Med. Biol. 23, 159 (1978).
16. A. G. Tkachenko, H. Xie, Y. Liu, D. Coleman, J. Ryan, W. R. Glomm, M. K. Shipton, S. Franzen, D. L. Feldheim, *Bioconjugate Chem. 15*, 482-490 (2004).
17. N. L. Rosi, C. A. Mirkin. *Chem. Rev. 105*, 1547–1562 (2005).
18. B.N. Khlebtsov, N.G. Khlebtsov, Journal of Quantitative Spectroscopy & Radiative Transfer. 106, 154 (2007).
19. L. Novotny, B. Hecht, Principle of Nano-optics (Cambridge University Press, 2006); D. Sarid, W. A. Challener, Modern Introduction to Surface Plasmons: Theory, Mathematica Modeling, and Applications (Cambridge University Press, 2010).
20. M. R. Singh, Electronic, Photonic, Polaritonic and plasmonic Materials. Toronto: Wiley Custom (2014).

21. M. Singh, J. Cox, M. Brzozowski, J. Phys. D: Applied Physics. 47, 085101(2014); M. Singh, K. Davideu, J. Carson, J. Phys. D: Applied Physics. 49, 445103 (2016).
22. M. O. Scully, M. S. Zubairy, Quantum Optics (Cambridge University Press, 1997).
23. G. S. Agarwal, Springer Tracts in Modern Physics. 70, 1-128 (1976).
24. O. G. Calderón, M. A. Antón, and F. Carreño, Eur. Phys. J. D. 25, 77 (2003).